\documentclass[aps,nofootinbib,superscriptaddress,twocolumn]{revtex4-1}
\usepackage{graphicx}
\usepackage{amsmath}
\usepackage{hyperref}
\usepackage{amssymb,latexsym,mathrsfs}

\hypersetup{
    colorlinks=true,
    linkcolor=red,
    citecolor=blue,
}

\def\be{\begin{equation}}
\def\ee{\end{equation}}
\def\ba{\begin{eqnarray}}
\def\ea{\end{eqnarray}}

\def\l{\left}
\def\r{\right}

\begin{document}

\title{Observable physical modes of modified gravity}

\date{\today}

\author{Alireza Hojjati}
\affiliation{Department of Physics, Simon Fraser University, Burnaby, BC, V5A 1S6, Canada}
\affiliation{Department of Physics and Astronomy, University of British Columbia, Vancouver, BC, V6T 1Z1, Canada}
\author{Levon Pogosian}
\affiliation{Department of Physics, Simon Fraser University, Burnaby, BC, V5A 1S6, Canada}
\author{Alessandra Silvestri}
\affiliation{SISSA - International School for Advanced Studies, Via Bonomea 265, 34136, Trieste, Italy}
\affiliation{INFN, Sezione di Trieste, Via Valerio 2, I-34127 Trieste, Italy}
\affiliation{INAF-Osservatorio Astronomico di Trieste, Via G.B. Tiepolo 11, I-34131 Trieste, Italy}
\author{Gong-Bo Zhao}
\affiliation{National Astronomy Observatories, Chinese Academy of Science, Beijing, 100012, P.R.China}
\affiliation{Institute of Cosmology and Gravitation, University of Portsmouth, Portsmouth, PO1 3FX, UK}

\begin{abstract}
At linear order in cosmological perturbations, departures from the growth in the cosmological standard model can be quantified in terms of two functions of redshift $z$ and Fourier number $k$. Previous studies have performed principal component forecasts for several choices of these two functions, based on expected capabilities of upcoming large structure surveys. It is typically found that there will be many well-constrained degrees of freedom. However, not all and, probably most, of these degrees of freedom were physical if the parametrization had allowed for an arbitrary $k$-dependence. In this paper, we restrict the $k$-dependence to that allowed in local theories of gravity under the quasi-static approximation, i.e. ratios of polynomials in $k$, and identify the best constrained features in the ($z$,$k$)-dependence of the commonly considered functions $\mu$ and $\gamma$ as measured by an LSST-like weak lensing survey.  We estimate the uncertainty in the measurements of the eigenmodes of modified growth. We find that imposing the theoretical prior on $k$-dependence reduces the number of degrees of freedom and the covariance between parameters. On the other hand, imaging surveys like LSST are not as sensitive to the $z$-dependence as they are to the $k$-dependence of the modified growth functions. This trade off provides us with, more or less, the same number of well-constrained eigenmodes (with respect to our prior) as found before, but now these modes are physical.
\end{abstract}

\maketitle

\section{Introduction}
\label{sec1}

While the cause of cosmic acceleration remains a mystery for modern cosmology,  it has become increasingly clear that probes of cosmic structure will play a key role in discerning among candidate models. Upcoming and future redshift and weak lensing (WL) surveys, such as Dark Energy Survey (DES)~\cite{des}, Large Synoptic Survey (LSST)~\cite{lsst,Ivezic:2008fe} and Euclid~\cite{euclid}, will map the evolution of matter and metric perturbations from the matter dominated epoch until today with a high degree of precision. Combined with the measurements of the cosmic microwave background (CMB) by Planck~\cite{planck}, they will test the relations between the matter distribution, the gravitational and the lensing potentials on cosmological scales, thus shedding light on the underlying theory of gravity. 

In recent years there has been a growing effort to conceive optimal ways of testing General Relativity (GR) on cosmological scales in a model independent way~\cite{Linder:2007hg,Zhang:2007nk,Hu:2007pj,Daniel:2008et,Song:2008vm,Skordis:2008vt,Song:2008xd,Zhao:2009fn,Zhao:2010dz,Song:2010rm,Daniel:2010ky,Pogosian:2010tj,Bean:2010zq,Thomas:2011pj,Baker:2011jy,Thomas:2011sf,Zhao:2011te,Hojjati:2011xd,Bertschinger:2011kk,Sawicki:2012re,Baker:2012zs,Amendola:2012ky,Hojjati:2012ci,Motta:2013cwa,Asaba:2013xql,Baker:2013hia,Dossett:2013npa}. Since we are dealing with perturbations around a time dependent background, it is generally the case that deviations from the standard $\Lambda$ Cold Dark Matter ($\Lambda$CDM) cosmological model are parametrized in terms of unknown functions of time and scale. 

For a given expansion history, in addition to conservation of energy-momentum, two more equations are needed to close the system of linear order equations for scalar cosmological perturbations.  These equations that are parametrized with two functions, commonly chosen to be $\mu(a,k)$ and $\gamma(a,k)$ (see Eq.~(\ref{eq:gamma_mu_def}) in Sec.~\ref{sec2}), can reproduce solutions for scalar modes in all possible modifications of gravity as well as models with exotic dark fluids~\cite{Pogosian:2010tj,Bean:2010zq,Bertschinger:2006aw, Amendola:2007rr,Bertschinger:2008zb}. While these functions, which we will refer to as the modified growth (MG) functions, represent a practical and theoretically consistent parametrization of deviations from $\Lambda$CDM, one needs to make further assumptions about their form when performing forecasts or fits to data. With the aim of not being biased by specific theoretical models, we have previously performed a principal component analysis (PCA) of $\mu$ and $\gamma$ treating them as arbitrary functions of scale and time~\cite{Zhao:2009fn,Hojjati:2011xd,Asaba:2013xql}. However, an arbitrary $k$-dependence allows for unphysical possibilities -- in fact, most of the well-constrained eigenmodes are probably not physical. The scale-dependence of the terms in the equations of motion and of MG functions has been investigated for a variety of modified gravity models in~\cite{Baker:2012zs,DeFelice:2011hq,Amendola:2012ky,Brax:2004qh,Brax:2005ew,Brax:2013mua}. More generally, in~\cite{Silvestri:2013ne} it was shown that, in \emph{local} theories of gravity, under the quasi-static approximation, MG functions must be ratios of polynomials in the wavenumber; furthermore, the polynomials are even and of second degree in practically all viable models of cosmic acceleration considered today. In this paper, we translate such theoretical considerations into a practical prior on the  $k$-dependence of MG functions in order to identify the modes that are physical. As we will show, this significantly reduces the uncertainty associated with the physically interesting modes.

Other physically consistent parameterizations of cosmological perturbations have been developed in the literature, {\it e.g.} one inspired by the effective field theory (EFT) approach~\cite{Gubitosi:2012hu,Bloomfield:2012ff,Piazza:2013coa}. There, one can describe the evolution of the background and the dynamics of linear scalar perturbations in all single scalar field dark energy and modified gravity models with only six functions of time~\cite{Gleyzes:2013ooa,Bloomfield:2013efa}. The advantage of the EFT parametrization is that these functions are linked to the terms appearing in a Lagrangian constructed according to precise and definite rules. Specifically, they multiply all terms that are consistent with the unbroken symmetries of the theory in the unitary gauge ({\it i.e.} spatial diffeomorphisms), and are at most quadratic in perturbations. While, in principle, one could link each of these terms, and therefore each of the functions multiplying them,  to a corresponding observable effect, in practice, such an identification is not always feasible; in this formalism, quantities that are more directly linked to observables, such as the effective Newton constant and the gravitational slip, correspond to complicated expressions in terms of the time dependent coefficients of all terms in the action. Similar issues are faced by the ``new PPF'' parametrization of~\cite{Baker:2012zs} and the ``equations of state'' approach of~\cite{Battye:2012eu,Battye:2013aaa,Battye:2013ida}. In our view, for the purpose of extracting and storing information from data about deviations from the $\Lambda$CDM model, the MG functions represent a more practical option.

In this paper, we estimate the number of degrees of freedom of $\mu$ and $\gamma$, with the physically allowed $k$-dependence, that can be measured by an LSST-like survey. We also address how well one can detect the existence of $k$-dependence in MG functions. We also pay special attention to the detectability of $\gamma \ne 1$, since it would be a signature of a new gravitational degree of freedom~\cite{Kunz:2006ca,Daniel:2008et,Amendola:2013qna}.

The paper is organized as follows. In Sec.~\ref{sec2}, we briefly review common choices of MG functions,  their respective pros and cons and the theoretical considerations on their scale dependence. We then focus on the pair $(\mu,\gamma)$. In Sec.~\ref{sec3}, we review the results from PCA of MG functions from previous work~\cite{Zhao:2009fn,Hojjati:2011xd} and explain the implementation of the theoretical prior on their $k$-dependence, which reduces the problem to a PCA of \emph{five} functions of time. As we show in Sec.~\ref{sec4}, these functions are highly degenerate among themselves and hence cannot be individually constrained. We therefore consider the eigenmodes of the combinations (triplets) of these functions that enter each of the MG functions ($\mu$ and $\gamma$) and use them to ``assemble'' the corresponding $\mu$, $\gamma$ surfaces in the $(z,k)$ plane. 

We find that an LSST-like survey will only marginally constrain the $k$-dependence of the MG functions after they are restricted to the physical ansatz. Overall, about $10$ modes of the MG  functions can be constrained with uncertainties smaller than $1$\% of the prior (see Section \ref{sec3}) when one considers the combined effect of the two functions. Individual constraints on $\mu$ and $\gamma$, after accounting for the covariance between them, are expected to be weak, with $\mu$ being the better constrained function. 

\section{Modified growth functions}
\label{sec2}

\subsection{($\mu$,$\gamma$) vs (${\mu}$,$\Sigma$)}
The evolution of scalar linear perturbations can be fully described by a choice of two MG functions that relate the Newtonian metric potential to the matter density contrast and to the curvature potential~\cite{Pogosian:2010tj,Bean:2010zq,Bertschinger:2006aw,Amendola:2007rr,Bertschinger:2008zb}. They are defined in Fourier space via
\be\label{eq:gamma_mu_def}
k^2\Psi=-4\pi Ga^2\mu(a,k)\rho\Delta\,,\,\,\,\,\,\Phi=\gamma(a,k)\Psi \ .
\ee
In the above, $a$ is the scale factor, $\rho$ is the background matter density, $\Delta$ is the comoving matter density contrast, and $\Psi$ and $\Phi$ are the scalar metric potentials in conformal gauge, describing fluctuations in the Newtonian potential and the curvature, respectively. Unless explicitly stated otherwise, we work with Fourier transforms of all cosmological perturbations. The algebraic relations~(\ref{eq:gamma_mu_def}) are combined with the equations for the conservation of the matter energy-momentum to form a complete set of equations. 
 
An alternative way to close the system of equations is to introduce $\Sigma$ alongside $\mu$, namely use
\ba \nonumber
k^2\Psi=-4\pi Ga^2{\mu}(a,k)\rho\Delta , \\
k^2\l(\Phi+\Psi\r)=-8\pi G a^2\Sigma(a,k)\rho\Delta \ .
\label{eq:sigma_def}
\ea

It is important to realize that $\mu$ that is part of the $(\mu,\gamma)$ parametrization can have a different impact on the observables ({\it e.g.} WL) from when it is part of the $({\mu},\Sigma)$ pair. To see this, imagine varying $\mu$ while holding $\Sigma$ fixed. The corresponding change in the WL potential, $\Phi+\Psi$, will be different from the case in which one varies $\mu$ but holds $\gamma$ fixed. Thus, the observational error bars one obtains for $\mu$ depend on the choice of the second function with which it was co-varied. This argument obviously also applies to $\gamma$ and other cosmological parameters  -- the uncertainty in a given parameter depends on the choice of other parameters it was co-varied with.

Given that the choice of the two algebraic relations is not unique, one picks specific MG functions depending on the theoretical motivation and observables under consideration. It is common to use $\mu$ because it is the only function entering in the evolution equation for $\Delta$ in the quasi-static limit and, therefore, is directly probed by galaxy counts (GC) and other tracers of the Newtonian potential. $\gamma$ is a direct indicator of a non-minimal coupling between the matter and the metric. Indeed, while $\mu \ne 1$ can be due to a minimally coupled DE or massive neutrinos, $\gamma \ne 1$ would imply a change in the way that massive particles respond to the metric~\cite{Kunz:2006ca,Daniel:2008et,Amendola:2013qna}, {\it i.e.} constitute a modification of gravity. Another advantage of working with the $(\mu,\gamma)$ pair is that the corresponding system of equations has a clear super-horizon limit in which, as expected, $\gamma$ is the only MG function playing a role~\cite{Pogosian:2010tj}. On the other hand, the $({\mu},\Sigma)$ pair is more directly measured by WL~\cite{Zhao:2010dz,Song:2010fg}, in the sense that it is easier to break the degeneracy between ${\mu}$ and $\Sigma$ than between $\mu$ and $\gamma$. However, $\Sigma$ is less interesting from the theoretical point of view, since its departures from unity are predicted to be negligible in all viable scalar-tensor theories studied in the literature. In other words, $\Sigma=1$ is more likely to be consistent with MG than $\gamma=1$. For these reasons, we choose to work with $(\mu,\gamma)$.

\subsection{The physical $k$-dependence}

A detailed PCA of MG functions has been performed in~\cite{Zhao:2009fn,Hojjati:2011xd,Asaba:2013xql} (see also~\cite{Hall:2012wd}) where $\mu$ and $\gamma$ were treated as arbitrary functions of $k$ and $z$. We will review some of these results in the next Section, but it is generally found that upcoming surveys will be able to constrain quite a few eigenmodes of the MG functions. However, it was not clear how many of these constrainable eigenmodes are of actual physical interest. An arbitrary relation between two quantities in Fourier space, such as those in Eq.~(\ref{eq:gamma_mu_def}), does not, in general, imply a local relation between them or their derivatives in real space. Therefore, the $k$-dependence of $\mu(a,k)$ and $\gamma(a,k)$ cannot be completely arbitrary if equations of motion are obtained from the variational principle. As shown in~\cite{Silvestri:2013ne}, based on considerations of locality and general covariance, and under the quasi-static approximation, physically acceptable forms of $\mu(a,k)$ and $\gamma(a,k)$ correspond to ratios of polynomials in $k$, with the numerator of $\mu$ set by the denominator of $\gamma$.  Furthermore, in models with purely scalar extra degrees of freedom, the polynomials are even in $k$ and in most, if not all, of the viable models considered so far in the literature, they are of second degree. This includes all single (scalar) field dark energy and modified gravity models. These arguments bring us to the following irreducible forms for $\gamma(a,k)$ and $\mu(a,k)$
\begin{subequations}\label{gamma_mu_p}
\ba
\label{eq:gamma}
\gamma &=&  {p_1(a)+p_2(a) k^2 \over 1+p_3(a) k^2}\ , \\
\mu &=& {1 + p_3(a) k^2 \over p_4(a) + p_5(a) k^2} \ .
\label{eq:mu}
\ea
\end{subequations}
In this paper we restrict the scale dependence of MG functions to this form and investigate their best constrained, physically allowed modes.

The scale-dependence of the dynamics of perturbations has been analyzed in other works. In~\cite{Baker:2012zs} Baker et al. investigated the form of the equations of motion under the quasi-static limit in the Parametrized Friedmann framework; they found that for several modified gravity models,  the equations reduce to algebraic ones containing only even powers of the wavenumber. Other authors have investigated the form of MG functions in the quasi-static regime for scalar-tensor~\cite{Brax:2004qh,Brax:2005ew,Brax:2013mua} and Horndeski theories~\cite{DeFelice:2011hq,Amendola:2012ky,Horndeski:1974wa}. The derivation of~\cite{Silvestri:2013ne} is general and based on simple  arguments of locality and covariance, and applies, among others,  to all the former cases.

Before concluding this discussion, let us notice that, while parametrization~(\ref{gamma_mu_p}) is motivated by the quasi-static regime, it allows for departures from $\Lambda$CDM on near-horizon scales as discussed in~\cite{Silvestri:2013ne}. 

\section{Implementing the theoretical prior on scale-dependence}
\label{sec3}

Applying the theoretical prior~(\ref{gamma_mu_p}) on the scale-dependence of $\gamma$ and $\mu$ reduces the problem of constraining two functions of two variables to that of five functions of one variable, {\it i.e.} the functions $p_\alpha(a)$. In this paper we {\em forecast} the constraints expected from a WL survey like LSST in combination with CMB data from Planck and supernovae from a future survey. Rather than calculating the Fisher matrices for the functions $p_\alpha(a)$ from scratch, we can derive them via a projection from the Fisher matrices for $\mu$ and $\gamma$ binned in $k$ and $z$ that have already been calculated in~\cite{Hojjati:2011xd}. Before discussing the details of this projection, let us briefly review the main results of~\cite{Hojjati:2011xd} where a PCA was performed for general $(\mu,\gamma)$. These results can then be compared with the ``physical'' modes of $(\mu,\gamma)$ that we will assemble from the eigenmodes of $p_\alpha$ in Sec.~\ref{sec4}.

\subsection{Forecasts for ($\mu$,$\gamma$) with arbitrary $k$ dependence}
\label{old_PCA}

\begin{figure*}[!tbh]
\includegraphics[width=0.9\columnwidth]{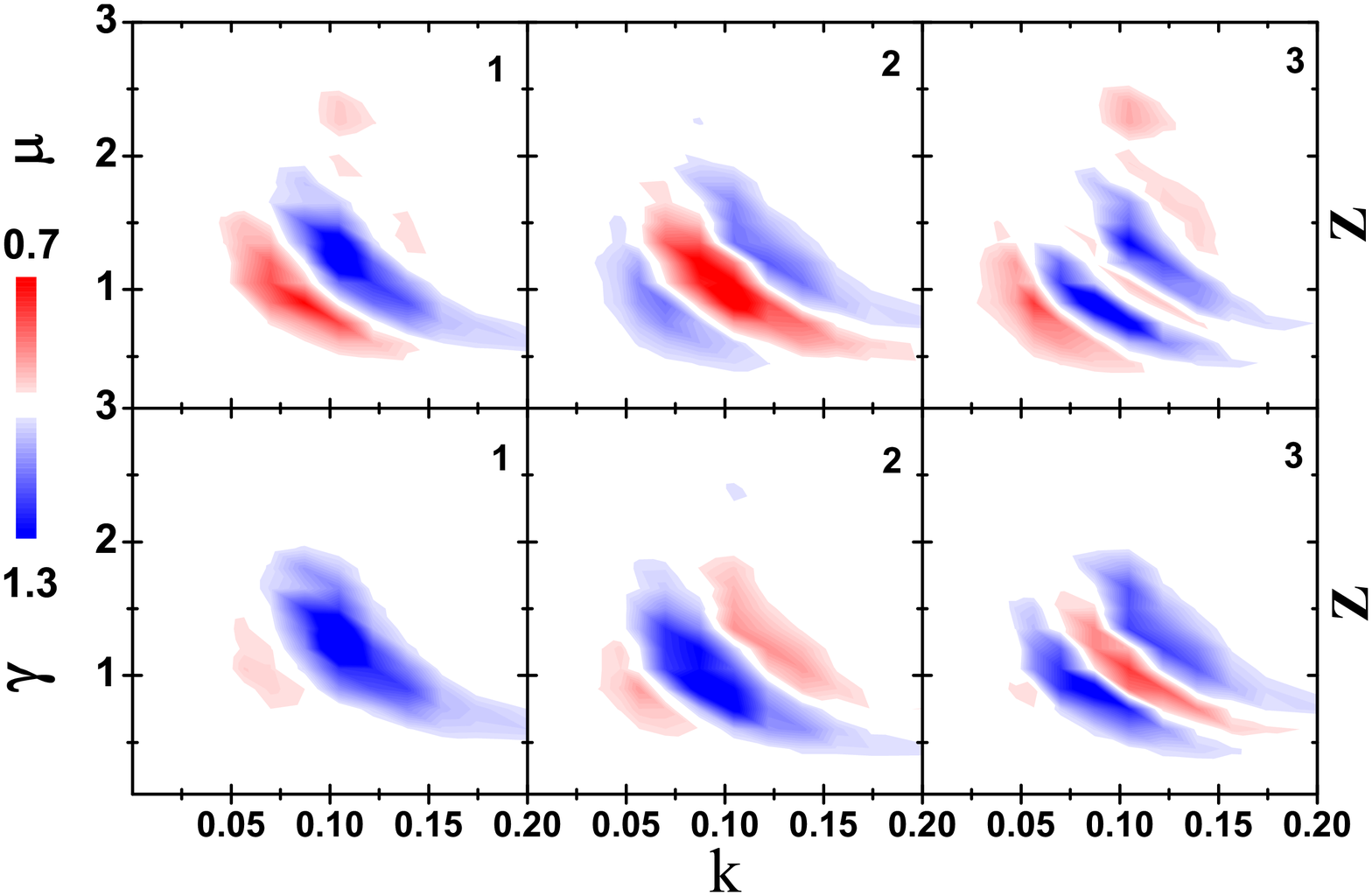}
\includegraphics[width=0.9\columnwidth]{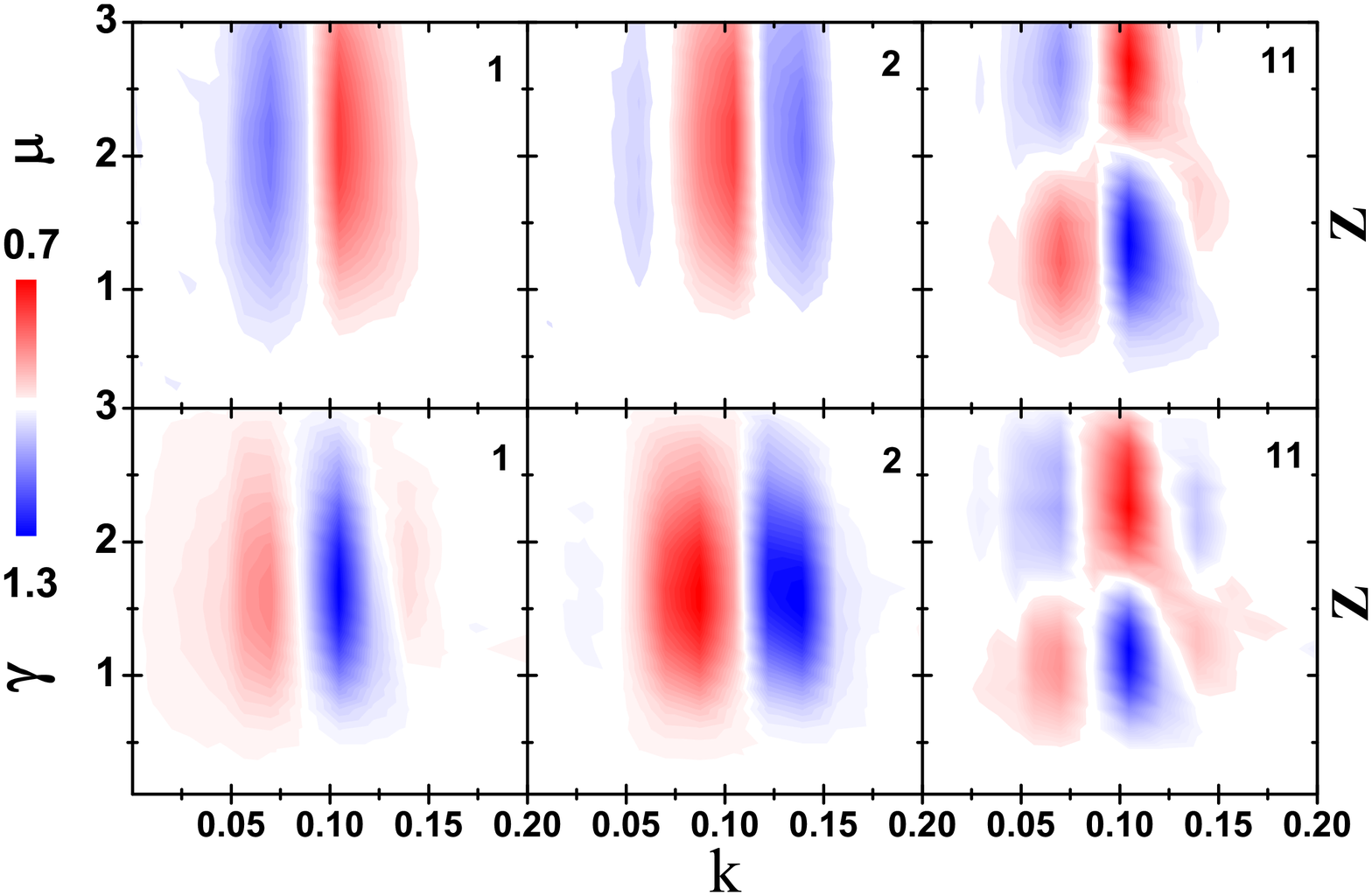}
\caption{{\em Left:} the first three best constrained eigenmodes of $\mu$ and $\gamma$ in the case when the two functions are not marginalized over each other. {\em Right:} the $1^{\rm st}$, $2^{\rm nd}$ and $11^{\rm th}$ eigenmodes of $\mu$ and $\gamma$ obtained after marginalizing over each other.}
\label{fig:mu-gamma}
\end{figure*}

\begin{figure}[tbh]
\includegraphics[width=0.9\columnwidth]{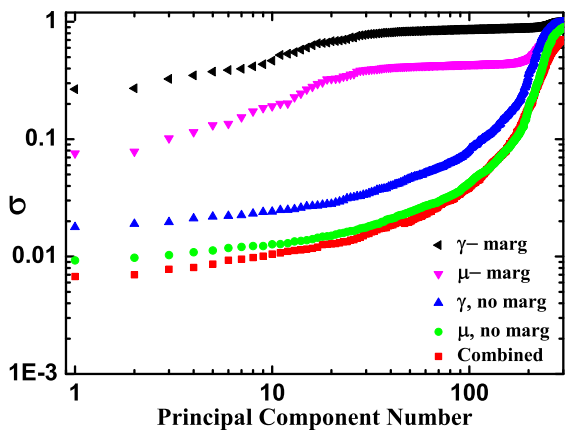}
\caption{Uncertainties associated with the eigenmodes of $\mu$ and $\gamma$ for three cases: unmarginalized over each other (with the $1^{\rm st}$, $2^{\rm nd}$ and $3^{\rm th}$ shown in the left panel of Fig.~\ref{fig:mu-gamma}), marginalized over each other  (with the $1^{\rm st}$, $2^{\rm nd}$ and $11^{\rm th}$ shown in the right panel of Fig.~\ref{fig:mu-gamma}), and combined. }
\label{fig:mu-gamma_errors}
\end{figure}

In~\cite{Hojjati:2011xd} we have performed a PCA of ($\mu$,$\gamma$) treating them as general functions of $k$ and $z$. In that approach, $\mu$ and $\gamma$ were discretized into ``pixels'' in ($z,k$) space, with 20 bins in k in the linear regime, and 20 bins in z, uniformly spaced between 0 and 3, along with a ``high-z'' bin between 3 and 30. We then calculated the Fisher matrix for $878$ parameters, which included $840$ pixels of ($\mu$,$\gamma$). The other 38 parameters were the main cosmological parameters, including dark energy equation of state $w(z)$ binned on the same grid in $z$, galaxy bias and supernova nuisance parameter (note that throughout this paper, unless ecplicitly mentioned, we always marginalize over these 38 parameters). In that forecast, we have considered WL and galaxy counts (GC) spectra, as well as the WL-GC cross-correlation spectra, from an LSST-like survey, combined with temperature and polarization spectra from Planck and $4000$ supernovae in 14 redshift bins between $z=0.15$ and $1.5$, as can be expected from a future space based SNIa survey.

Note that while the highest redshift pixels ($3<z<30$) of $\mu$ and $\gamma$ are outside the range directly probed by the WL surveys, they do impact the observables: $\mu(z>3,k)$ re-sets the amplitude of the growth at all lower redshifts, while $\gamma(z>3,k)$ affects the Integrated Sachs-Wolfe (ISW) contribution to the CMB temperature anisotropy. The sensitivity of observables to variations in $\mu(z>3,k)$ and $\gamma(z>3,k)$ depends on the assumed high-$z$ cutoff -- making the high-z interval wider increases the sensitivity. These variations are also degenerate with other parameters that affect the overall growth between recombination and $z=3$ (see~\cite{Hojjati:2011xd,Hojjati:2012ci} for a detailed discussion of this and other degeneracies). Thus it makes sense to focus on quantities that are independent of this cutoff, such as bounds on $\mu$ and $\gamma$ obtained after marginalizing over the high-z MG pixels, as well as all other cosmological parameters.

Since the individual pixels of $\mu$ and $\gamma$ are highly correlated, their uncertainties in any particular bin will be practically unconstrained. The PCA is a way to de-correlate the parameters and find their linear combinations -- the {\em eigenmodes} -- that are best constrained by data. We refer the reader to~\cite{Hojjati:2011xd} for a detailed review of the PCA approach to modified growth. We show a few representative eigenmodes of $\mu$ and $\gamma$ in Fig.~\ref{fig:mu-gamma}. The left panel shows the ``unmarginalized'' eigenmodes of $\mu$ and $\gamma$, {\it i.e.} the modes of each function obtained after marginalizing over the cosmological parameters but not over each other. This means that we show the eigenmodes of $\mu$ with $\gamma$ fixed to its fiducial value of $1$, and eigenmodes of $\gamma$ with $\mu$ fixed at 1. The right panel shows the marginalized eigenmodes, where one now sees the eigenmodes of $\mu$ after taking into account its covariance with $\gamma$ (and vice versa). The corresponding uncertainties are plotted in Fig.~\ref{fig:mu-gamma_errors}. In the marginalized case, one effectively throws away all information that is unable to distinguish between the two functions. For instance, if one wanted to know specifically if $\gamma \ne 1$, which is an important test of MG, one would want to exclude from consideration the effects of $\mu \ne 1$ that can be confused with variations in $\gamma$. As one can see from the large uncertainties on marginalized $\gamma$ modes in Fig.~\ref{fig:mu-gamma_errors}, the degeneracy with $\mu$ significantly obscures measurements of $\gamma$.

A few observations can be made from Fig.~\ref{fig:mu-gamma} that will be relevant for the discussion that follows. In the unmarginalized case (the left panel of Fig.~\ref{fig:mu-gamma}), the eigenmodes have a diagonal form in the $(z,k)$ space, showing a degeneracy between scale and time. This is because the WL observables dominate the information provided by the experiments we consider. Indeed,  the changes in the weak lensing kernel due to a shift of the lens along the line of sight ({\it i.e.} a change in redshift) are degenerate with those due to a resizing of the lens ({\it i.e.} a change in scale, or $k$). In the marginalized case, this degeneracy between variations in $k$ and $z$ disappears because marginalization erases all signatures that are common to both functions.

If we look at the shapes of the first few best constrained eigenmodes in the right panel of Fig.~\ref{fig:mu-gamma}, we note that they have nodes in $k$ direction, but not in $z$. It is only after the 11th best constrained mode of $\mu$ (and $\gamma$) that we start to see a node in $z$. This may lead one to conclude that the experiments we have considered are more sensitive to the $k$ dependence of MG functions than their time dependence. On one hand, this is expected -- after all varying $\mu$ and $\gamma$ at each $k$-bin is equivalent to allowing for arbitrary $k$ dependent normalization of the spectra that is easily detectable. On the other hand, we know that the $k$-dependence in these functions comes from spatial derivatives in the equations of motion and, hence, cannot be completely arbitrary~\cite{Silvestri:2013ne}. Hence, in what follows, we will examine to what extent the sensitivity to $k$ dependence diminishes after restricting the forms of the MG functions to those in~(\ref{gamma_mu_p}).

\subsection{Projection onto the $p$'s}\label{projection}

Given the analytical expressions~(\ref{gamma_mu_p}) for $\mu$ and $\gamma$, we can derive the Fisher matrix for binned values of $p_\alpha(a)$ from the Fisher matrix of the $(\mu,\gamma)$ pixels via projection of errors. Let us introduce parameters $m^{(a)}_{ij}$, such that $m^{(1)}_{ij}=\mu(a_i,k_j)$ and $m^{(2)}_{ij}=\gamma(a_i,k_j)$, and collectively label all other cosmological parameters as $\omega$. Let us also define the new parameters as $p_\alpha(a_\ell)$, where $\alpha=1,...,5$ labels the functions and $a_\ell$ labels the bins in $a$ taken to correspond to the original bins in $z$  used for discretizing $\mu$ and $\gamma$ (we work with $a(z=0)=1$). Then, the projected Fisher matrix elements  are
\ba\label{Fisher_ps}
\nonumber
F^P_{p_\alpha(a_\ell),p_\beta(a_\gamma)} &=& \sum_{aijbmn} {\partial m^a_{ij} \over \partial p_\alpha(a_\ell)} F_{m^a_{ij},m^b_{mn}} {\partial m^b_{mn} \over \partial p_\beta(a_\gamma)} \\
&=& \sum_{ajbn} {\partial m^a_{\ell j} \over \partial p_\alpha(a_\ell)} F_{m^a_{\ell j},m^b_{\gamma n}} {\partial m^b_{\gamma n} \over \partial p_\beta(a_\gamma)} \ ,
\ea
where the derivatives are evaluated at the fiducial values, {\it i.e.} $p_1=p_4=1$ and $p_2=p_3=p_5=0$, and the second line follows from derivatives being non-zero only if evaluated at the same bin in $a$. Similarly,
\be
F^P_{p_\alpha(a_\ell),\omega} = \sum_{ja} {\partial m^a_{\ell j} \over \partial p_\alpha(a_\ell)} F_{m^a_{\ell j},\omega} \ {\rm and} \ 
F^P_{\omega \omega}= F_{\omega \omega} \ ,
\ee
with the partial derivatives given by 
\ba
 &&{\partial m^1_{ij} \over \partial p_3(a_\ell)} =\delta_{i\ell}k_j^2,\,\, {\partial m^1_{ij} \over \partial p_4(a_\ell)} = -\delta_{i\ell},\,\,
 {\partial m^1_{ij} \over \partial p_5(a_\ell)} =- \delta_{i\ell} k_j^2, \nonumber\\
&& {\partial m^2_{ij} \over \partial p_1(a_\ell)} = \delta_{i\ell},\,\,{\partial m^2_{ij} \over \partial p_2(a_\ell)} =\delta_{i\ell}k^2_j,\,\,  {\partial m^2_{ij} \over \partial p_5(a_\ell)} = - \delta_{i\ell} k_j^2,  \nonumber \\
 && {\partial m^1_{ij} \over \partial p_1(a_\ell)} = {\partial m^1_{ij} \over \partial p_2(a_\ell)} = 
 {\partial m^2_{ij} \over \partial p_4(a_\ell)} =  {\partial m^2_{ij} \over \partial p_5(a_\ell)} =0 \ .
\ea

For sufficiently fine binning, the projected constraints should not depend on the size of the bins in $k$. Finer binning adds terms to the Fisher matrix, but the error on the corresponding pixels is larger because observables are less sensitive to a narrower bin. In the limit of many bins, both effects scale proportionally to the width of the bin and cancel out in the Fisher matrix.

We choose to work with $k$ in units of $k_*=0.2$ h/Mpc which makes all functions $p_\alpha$ dimensionless. It also assures that a simple Gaussian prior on $\mu$ and $\gamma$ in each pixel, that approximately restricts them to the $1 \pm \sigma_P$ range, implies a restriction on the values of $p_\alpha$ to also lie in the $\pm \sigma_P$ vicinity of their fiducial values. 

Aside from being necessary to prevent singularities in the Fisher matrix inversion, priors allow for a meaningful interpretation of the PCA results. Namely, while some eigenmodes are better constrained than others, without a prior all eigenmodes are technically informative. On the other hand, we can state how many eigenmodes can be measured with uncertainties that improves on the given prior by a certain amount. The only meaningful way to justify the value of any prior is to appeal to theoretical expectations. Based on our knowledge of MG models studied in the literature, we conservatively set $\sigma_P=1$. While we adopt a simple prior in which pixels are assumed to be independent, we note that a more realistic prior would be one in which they are taken to be correlated~\cite{Crittenden:2005wj,Crittenden:2011aa}. This is important when attempting a reconstruction of the unknown functions from data~\cite{Crittenden:2011aa,Zhao:2012aw}, but not crucial in the context of this paper.

\section{Results}
\label{sec4}

 \begin{figure}[tbp]
\includegraphics[width=0.85\columnwidth]{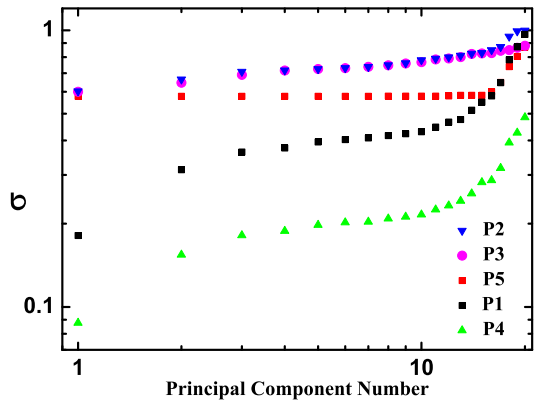}
\caption{Uncertainties (square roots of the covariance matrix eigenvalues) associated with the eigenmodes of the individual $p_\alpha$ functions. For each function, these account for the covariance with the other four functions.}
\label{fig:Errors-individual-marginalized}
\end{figure} 

 \begin{figure}[tbp]
\includegraphics[width=0.85\columnwidth]{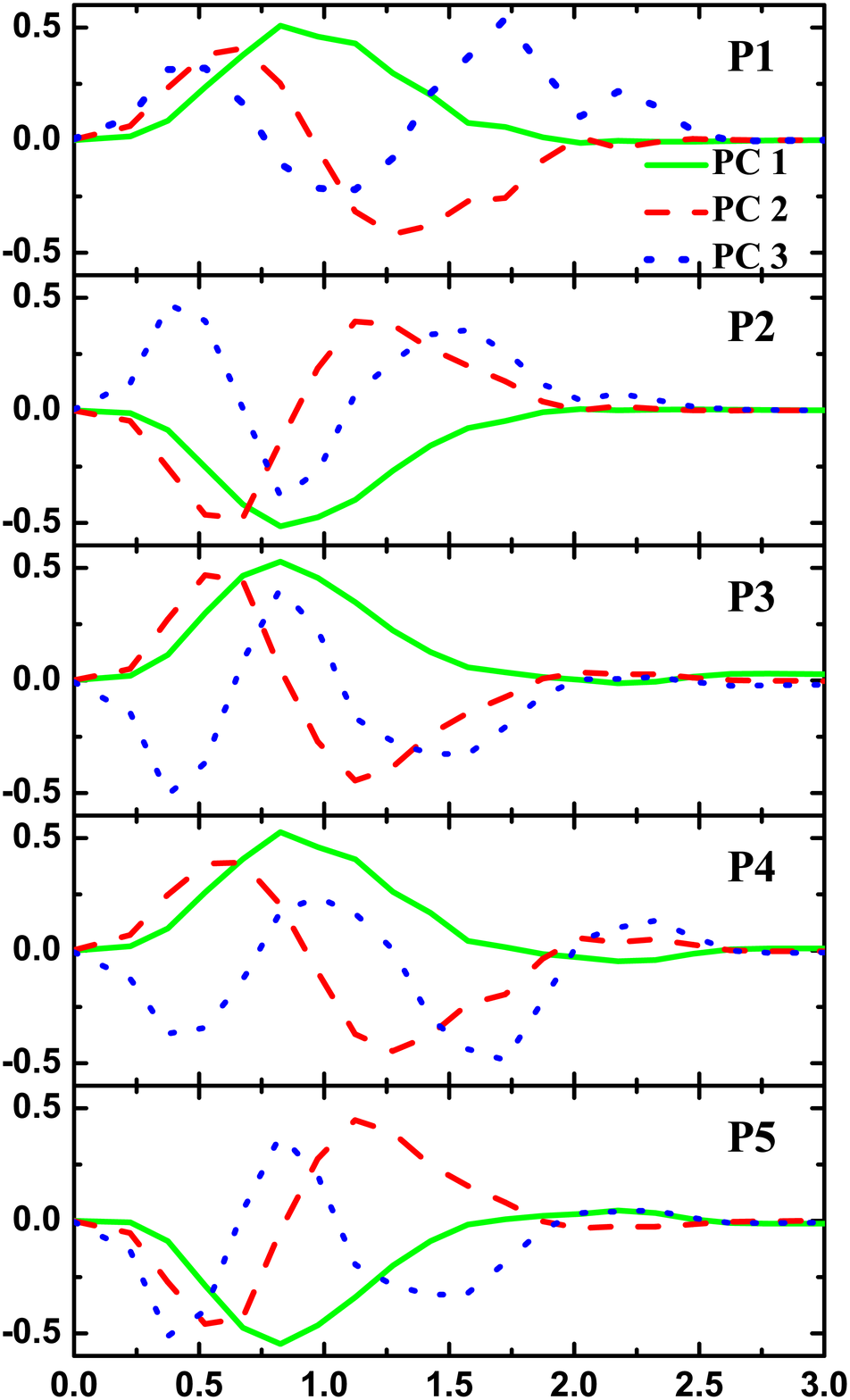}
\caption{The first three best constrained eigenmodes for each of the functions $p_\alpha$. For each function, these are obtained after marginalizing over the other four functions.}
\label{fig:eigenmodes-individual_marginalized}
\end{figure} 

 \begin{figure}[tbp]
\includegraphics[width=0.9\columnwidth]{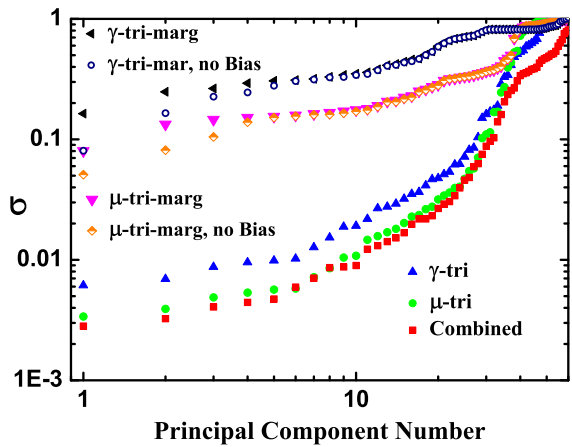}
\caption{Uncertainties associated with the eigenmodes of $\mu$ and $\gamma$ triplets for two cases: unmarginalized and marginalized over the other two functions ($p_3$ is common to both triplets and is not marginalized over).  The hollow symbols correspond to the case in which the galaxy bias is assumed to be known. The uncertainties on the combined eigenmodes of the 5 functions is also shown (red squared).}
\label{fig:errors-triplets-full}
\end{figure}

\begin{figure*}[!tbh]
\includegraphics[width=0.9\columnwidth]{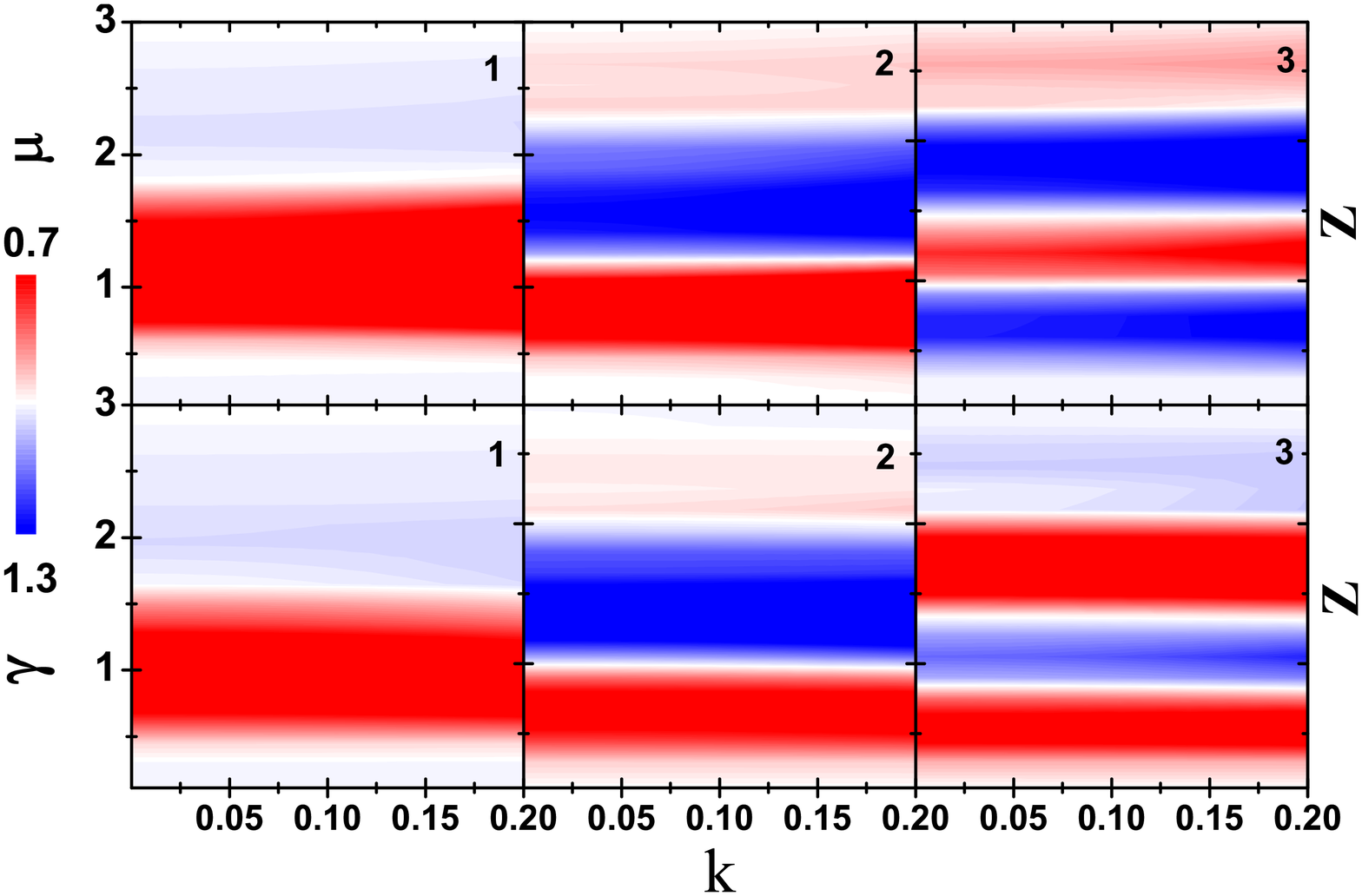}
\includegraphics[width=0.9\columnwidth]{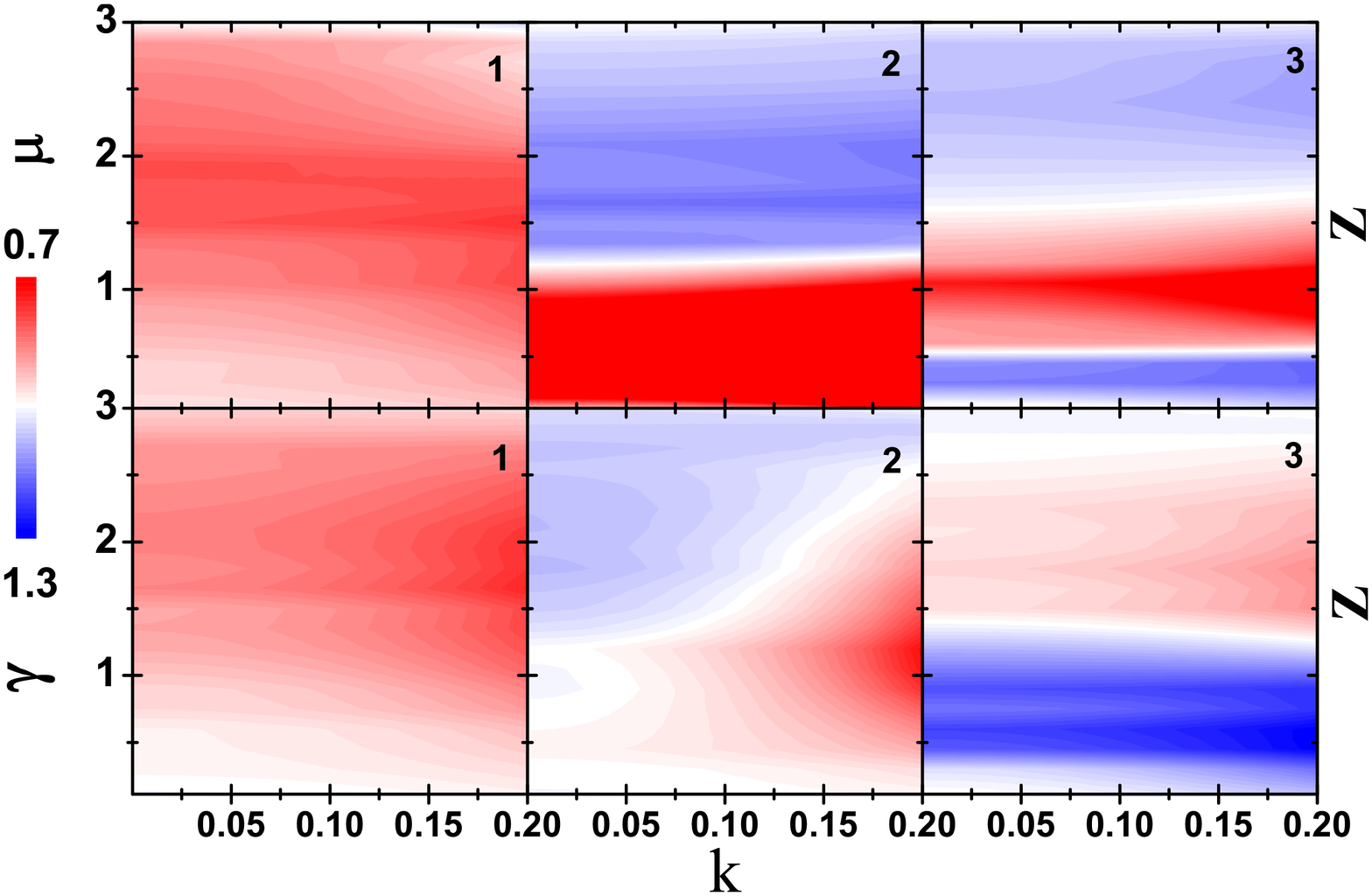}
\caption{ $\mu(a,k)$~(\ref{eq:mu}) and $\gamma(a,k)$~(\ref{eq:gamma}) assembled from the first three best constrained eigenmodes of their triplets, {\it i.e.} $\{p_3,p_4,p_5\}$ and $\{p_1,p_2,p_3\}$ respectively. The \emph{left panel} shows the unmarginalized case, while the \emph{right} panel shows the marginalized case.}
\label{fig:reconstructed1}
\end{figure*}

With the framework set up in the previous sections, we now proceed to answer the following questions:
\begin{itemize} 
\item How well can one measure functions $p_{\alpha}$ individually? 
\item How well can one detect departure of any of the five functions from the fiducial value without trying to identify them individually? This is the same as asking how well can one detect departures of either $\mu$ or $\gamma$ from their fiducial values, after constraining their $k$-dependence to ratios of polynomials; 
\item How well can one detect departures specific to $\mu$ or $\gamma$, especially the latter, since it is an important trigger of MG? 
\item What types of patterns in the ($z,k$) dependence of $\mu$ and $\gamma$ are best constrained?
\item How well can one detect a signature of $k$-dependence?
\end{itemize}

We can answer the above questions by starting from the projected Fisher matrix~(\ref{Fisher_ps}) and inverting different blocks. This gives the covariance matrix for the parameters that were included in the inversion. In this way we can determine covariance of individual $p_\alpha$ or their combinations.

\subsection{Individual $p_\alpha$}

To forecast constraints on each individual $p_\alpha$, we invert the full Fisher matrix~(\ref{Fisher_ps}) to obtain the covariance between all parameters. We then diagonalize the block of the covariance matrix that corresponds to the particular $p_\alpha$ of interest and find its principal components (eigenmodes) and the associated uncertainties. These uncertainties quantify how well one can measure the coefficients of the expansion of $p_\alpha$ into their eigenmodes. Fig.~\ref{fig:Errors-individual-marginalized} thus shows the uncertainties associated with the eigenmodes in the order from best to worst constrained -- the numbers on the $y$-axis are the square roots of the eigenvalues of the five covariance matrix blocks corresponding to the five $p_\alpha$. In Fig.~\ref{fig:eigenmodes-individual_marginalized} we also plot the few best constrained eigenmodes for each function.

The relative flatness of the uncertainty curves in Fig.~\ref{fig:Errors-individual-marginalized}, {\it i.e.} the small change between consecutive eigenvalues, suggests that $p_\alpha$ are not individually constrained. This was expected because of the way $p_\alpha$ enter the MG functions~(\ref{gamma_mu_p}).  Indeed, there is a strong degeneracy between the effects of the different functions on observables -- {\it e.g.} one can largely compensate for a variation in $p_2$ by a change in $p_3$. And this is in addition to the inherent degeneracy between $\mu$ and $\gamma$ discussed in Sec.~\ref{old_PCA}.

From Fig.~\ref{fig:Errors-individual-marginalized}  one can  also notice that the errors on $p_1$ and $p_4$ follow the same trend, with $p_4$ being better constrained as expected, since it is related to $\mu$ which is constrained better than $\gamma$ as discussed in~\cite{Hojjati:2011xd}. Comparing the errors forecasted for $p_1$ and $p_4$ to those for $p_2$, $p_3$ and $p_5$, we note that the latter functions have larger uncertainties which, technically, depend on the normalization of $k$. However, the meaningful way to determine if a parameter is well constrained is to consider the improvement on the prior. As mentioned in Sec.~\ref{projection}, we set the normalization of $k$ in a way that makes the prior on all $p_\alpha$ the same. We have also checked that changing the normalization of $k$ by orders of magnitude (and correspondingly adjusting the prior) results in the same improvement on the prior.

\subsection{Combinations of all $p_\alpha$ and those specific to $\mu$- and $\gamma$}
\label{triplets}

As we have shown in the previous subsection, functions $p_{\alpha}$ are unlikely to be individually constrained. However, this need not preclude us from gaining other, perhaps more important, information about MG. For instance, one may want to know how well it is possible to constrain any signature of MG, without trying to identify which of the five functions is responsible for it. In this case, the relevant {\em combined} eigenmodes are obtained by diagonalizing the block of the covariance matrix corresponding to all five $p_\alpha$. The eigenvalues give the associated uncertainties that are plotted in Fig.~\ref{fig:errors-triplets-full}. One can see from this plot that about $10$ eigenmodes can be measured with $\sigma \le 0.01$, which can be interpreted as being a better than $1$\% accuracy if we compare to the assumed prior of $\sigma_P=1$ (see Sec.~\ref{projection}). While this is about the same as the number of well constrained combined eigenmodes of $\mu$ and $\gamma$ with arbitrary $k$-dependence \cite{Hojjati:2011xd} (see Fig.~\ref{fig:mu-gamma_errors}), the best constrained modes in Fig.~\ref{fig:mu-gamma} feature oscillations in $k$ that are unlikely to be present in any realistic model of MG. Thus, our results imply a considerable improvement in the prospects of detecting the physically allowed MG patterns as a result of applying a prior on the $k$-dependence.

Another interesting question is the extent to which the functions $\mu$ and $\gamma$ can be constrained, since they correspond to theoretically interesting quantities in the quasi-static limit~\cite{Pogosian:2010tj,Daniel:2008et,Amendola:2013qna}. To address this, we can consider the combinations of $p_{\alpha}$ that appear in $\mu$ and $\gamma$ respectively, and look at their corresponding modes. Specifically, we consider combined modes of $\{p_1,p_2,p_3\}$ ($\gamma$ triplet) and combined modes of $\{p_3,p_4,p_5\}$ ($\mu$ triplet). Fig.~\ref{fig:errors-triplets-full} shows the uncertainties associated with the eigenmodes of these triplets for two cases. In one case, to which we refer to as {\em unmarginalized}, the triplet is varied, while the remaining two functions are not. In the other, which we call {\em marginalized}, all functions are varied and functions not belonging to the triplet under consideration are marginalized over. Note that $p_3$ appears in both of the triplets and we do not marginalize over it.

From Fig.~\ref{fig:errors-triplets-full}, we see that $\mu$ and $\gamma$ can be measured well only if varied one at a time. When varied together, our combination of observables can at best measure their eigenmodes with uncertainties on the order of $10\%$ of the prior, with $\mu$ being better constrained than $\gamma$. Note that the eigenvalues in the marginalized case are very much like those of $p_4$ and $p_1$ in Fig.~\ref{fig:Errors-individual-marginalized}, indicating that the sensitivity to the $k$-dependence in $\mu$ and $\gamma$ (set by functions $p_2$, $p_3$ and $p_5$) is weak when they are co-varied. We will return to the detectability of $k$-dependence in Sec.~\ref{kdep}.

It is interesting to check to what extent the weak constraints on $\gamma$ are caused by the uncertainty in the galaxy bias, as opposed to its intrinsic degeneracy with $\mu$. To this aim, in Fig.~\ref{fig:errors-triplets-full}, we also show the uncertainties associated with the eigenmodes of the $\mu$ and $\gamma$ triplets without marginalizing over bias. One can see that this considerably reduces the uncertainty in two modes of $\gamma$, although the uncertainty in the amplitude of the best constrained mode is still only constrained with about $8$\%  of the prior. This suggests that a future spectroscopic survey, such as Euclid, may perform somewhat better, but the degeneracy with $\mu$ will probably prevent a sub-percent measurement of $\gamma$. We plan to investigate this in a future publication.

\subsection{Assembling $\mu$ and $\gamma$ from $p_{\alpha}$}

\begin{figure}[tbp]
\includegraphics[width=0.9\columnwidth]{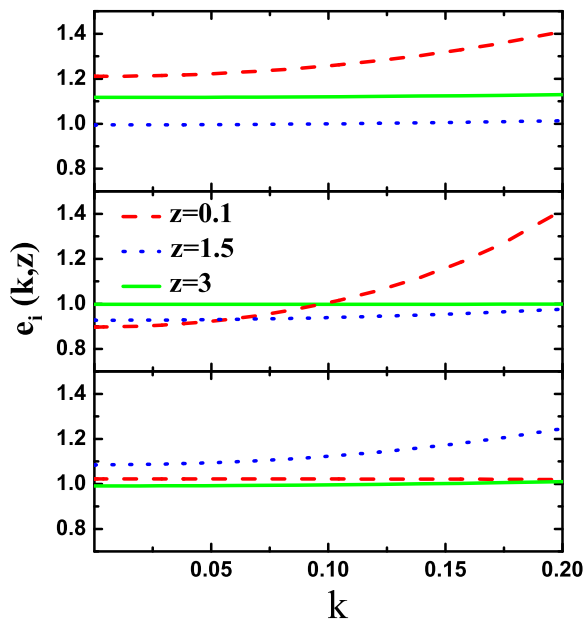}
\caption{We show the $k$-dependence in $\gamma$ assembled from the first three best constrained marginalized eigenmodes of the $\gamma$-triplet at three different redshifts.}
\label{fig:reconstructed2}
\end{figure}

\begin{figure}[tbp]
\includegraphics[width=0.9\columnwidth]{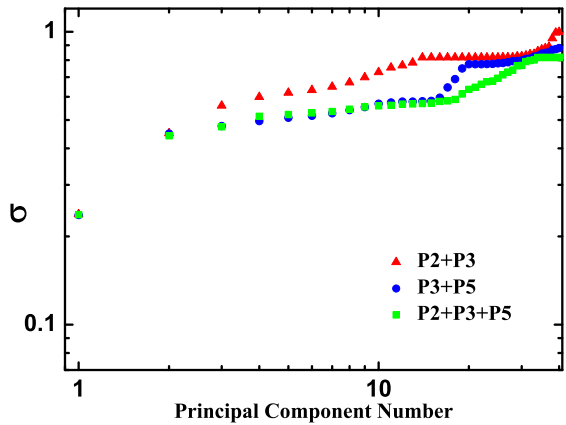}
\caption{Uncertainties associated with the eigenmodes of different combinations of $p_2$, $p_3$ and $p_5$. Red triangles show uncertainties on the pair $\{p_2,p_3\}$ which carries information on the $k$-dependence of $\gamma$, blue circles show uncertatinties on the pair $\{p_3,p_5\}$ which carries information on the $k$-dependence of $\mu$ and green squares show uncertainties on the triplet $\{p_2,p_3,p_5\}$ carrying information on any $k$-dependence in the evolution of linear perturbations.}
\label{fig:Errors-k-dep-marginalized}
\end{figure}
As mentioned above, $\mu$ and $\gamma$ are the functions that are interesting from the theoretical point of view. They are also more directly measured by the experiment than $p_\alpha$. Thus, in this subsection, we would like to see how the best constrained eigenmodes of $p_{\alpha}$ translate into features in the ($z,k$) dependence of $\mu$ and $\gamma$. This allows us to visualize the kind of scale- and time-dependent features that will be best constrained by the experiments after applying the theoretical prior~(\ref{gamma_mu_p}) on $k$-dependence. 

Fig.~\ref{fig:reconstructed1} shows the shapes of $\mu$ and $\gamma$ corresponding to the best three eigenmodes of their corresponding triplets, unmarginalized (left) and marginalized (right) over other functions. It is important to notice that these are not eigenmodes of $\mu$ and $\gamma$, {\it e.g.} they do not form an orthogonal basis. These are just the eigenmodes of the $p_{\alpha}$ substituted into Eqs.~(\ref{gamma_mu_p}) -- we can call them ``assemblies''.

A comparison of the unmargnalized with the marginalized assemblies in Fig.~\ref{fig:reconstructed1} shows that, analogous to what we saw for the eigenmodes of $\mu$ and $\gamma$ in Fig.~\ref{fig:mu-gamma}, marginalization smoothes out the features, which are primarily in the $z$-dependence in this case. We also see, from Fig.~\ref{fig:reconstructed2}, that the best constrained patterns in $\mu$ and $\gamma$ do not show significant $k$-dependence.

\subsection{The sensitivity to $k$-dependence}
\label{kdep}

One of the key signatures that distinguishes simple dark energy models, such as the minimally coupled scalar field quintessence models, from MG is the scale-dependence of the growth factor. Thus, the ability to detect the $k$-dependence in either $\mu$ or $\gamma$ is an important issue. Results of previous subsections indicate that removing the unphysical $k$-dependence from $\mu$ and $\gamma$ largely removes most of the sensitivity to their variations in $k$. It is worth recalling that this statement is specific to the collection of datasets that we consider in this paper, {\it i.e.} where information about MG comes primarily from WL and its cross-correlation with GC and we use specifications of LSST.

In Fig.~\ref{fig:reconstructed2}, we show the $k$-dependence in $\gamma$ corresponding to the first three best constrained eigenmodes of $\gamma$-triplets. For each triplet of eigenmodes, we plot the $k$-dependence of the assembled $\gamma$ at three different redshifts. With the functional form of $k$ fixed in our parametrization, the $k$-dependence in $\gamma$ comes from the departure of the functions $p_2$ and $p_3$ from their fiducial values. Also, the different shapes of the $k$-dependence at different redshifts are due to the time evolution of the functions in the triplet.

Clearly, some sensitivity to the $k$-dependence remains and we want to quantify it. For this, let us consider the functions $p_\alpha$ that multiply the $k^2$ factors, {\it i.e.} $p_2$, $p_3$ and $p_5$. We can then look at pairs of these functions corresponding to the scale dependence of $\gamma$ and $\mu$ separately, as well as the combination of all the three functions, which  carries information on any $k$-dependence in the linear growth. In Fig.~\ref{fig:Errors-k-dep-marginalized} we show the uncertainties associated with the eigenmodes of three different combinations: $\{p_2,p_3\}$, $\{p_3,p_5\}$, and $\{p_2,p_3,p_5\}$. The first combination carries the information about detectability of the $k$-dependence of $\gamma$, the second one on the $k$-dependence of $\mu$ and the last one on any $k$-dependence in the dynamics of linear perturbations.  None of them can be measured with an uncertainty smaller than $20$\% of the prior.

\section{Summary and conclusions}

In this paper we have investigated the best constrained features in the MG functions $\mu(z,k)$ and $\gamma(z,k)$ after restricting their scale-dependence to that allowed by local theories of gravity under the quasi-static approximation, i.e. ratios of polynomials in $k$. In particular, we focused on the case in which such polynomials are even and of second order, which includes most, if not all, of the viable models of dark energy and modified gravity considered so far in the literature. Applying this theoretical prior reduces the problem from that of constraining two functions of two variables to that of constraining five functions of time, such as $p_\alpha(a)$ in Eqs.~(\ref{gamma_mu_p}). Furthermore, it allows one to remove a priori from the analysis of modified growth many of the modes that are not physical; such modes were probably numerous among the eigenmodes of $\mu$ and $\gamma$ that we identified in previous PCA studies in which the $k$-dependence was unconstrained.

We focused on an LSST-like weak lensing survey and addressed several questions about the detectability of modified growth patterns both in terms of features in the time-dependence of $p_\alpha(a)$, individually and combined, and in terms of the original functions $\mu(z,k)$ and $\gamma(z,k)$ assembled from the corresponding triplets of $p_\alpha$.  As expected, we found that individual $p_\alpha$ are not constrained; indeed they are highly degenerate in their effects on observables because of the way they enter in the MG functions and because of the degeneracy between $\mu$ and $\gamma$ that they inherit. A more informative analysis is the one of detectability of combinations of $p_\alpha$. To this extent we considered different combinations: all five $p_\alpha$ to address the question of how well one can constrain any signature of modified growth without identifying which of the five functions is producing it; triplets corresponding respectively to $\mu$ and $\gamma$, since these are the functions that, in the quasi-static limit, offer the optimal bridge between theory and observations; pairs of the $p_\alpha$ corresponding respectively to the $k$-dependence in $\mu$ and $\gamma$ and, finally, the triplet of $p_\alpha$ carrying information on any $k$-dependence in the growth of structure. We have also substituted the eigenmodes of the $p_\alpha$ triplets appearing in $\mu$ and $\gamma$ into Eqs.~(\ref{gamma_mu_p}) and plotted the corresponding $\mu$ and $\gamma$ as functions of $(z,k)$.

 We find that about $10$ eigenmodes of the combination of all $p_\alpha$ will be measured  with uncertainty smaller that $1$\% of our prior, if one compares to an approximate prior of $\pm 1$ on the allowed deviation of $\mu$ and $\gamma$ from their $\Lambda$CDM values. If one wants to constrain combinations specific to $\mu$ and $\gamma$, the uncertainties are much larger, unless one fixes $\mu$ when varying $\gamma$ and vice versa. Our analysis also indicates that restricting to the physically allowed $k$-dependence of $\mu$ and $\gamma$ largely removes most of the sensitivity to their variations in $k$.

Detectability of $\gamma$ is of particular interest since a deviation of this function from its fiducial value in $\Lambda$CDM would signal a modification of the theory of gravity. We stress that such analysis needs to be performed with the function $\mu$ co-varied and marginalized over.  In our case, it is achieved by considering the eigenmodes of $\gamma$-triplets after marginalizing over $p_4$ and $p_5$, as we did in Sec.~\ref{triplets}.  We find that $\gamma$ is not expected to be well-constrained by LSST, with only one of the modes measured with a uncertainty smaller than $20$\%  of the prior. The $\mu$ triplets can be measured better, with one of the modes having an uncertainty below $10$\%. These seemingly pessimistic conclusions about detectability of $\gamma$ are specific to the collection of datasets that we considered in this paper, {\it i.e.} where information about MG comes primarily from WL and its cross-correlation with GC and we use specifications of LSST. We do expect that spectroscopic surveys will allow for an improvement in the detectability of MG functions, since they will provide a bias free measurement of the Newtonian potential $\Psi$. While we leave the full analysis for a spectroscopic survey, such as Euclid, for future work, we have checked what happens in the case of LSST when the bias is known, namely, if we do not marginalize over it. We find that fixing the bias resulted in a notable improvement in the uncertainty of the measurement of a couple of eigenmodes of $\gamma$-triplets. However, even with a perfect measurement of the WL shear and the galaxy redshifts, there will remain an irreducible uncertainty in measurements of $\gamma$ because of its degeneracy with $\mu$.

As a final remark, we note that, while PCA is a useful forecast tool, it does not tell us what parameters are best for fitting to data. One cannot simply fit only the best constrained modes, since this amounts to setting the amplitudes of the poorly constrained modes exactly to zero, which means one assumes to know their values with a perfect precision. Such an assumption can lead to a strongly biased reconstruction. Instead, it is possible to fit binned functions, supplemented by a prior on their smoothness, similarly to how it was done for $w(z)$ in~\cite{Zhao:2012aw}. Smoothness is a reasonable assumption for our five functions $p_\alpha$, since they are expected to be functions of the background only. We leave this analysis for future work.

\acknowledgments AH and LP are supported by NSERC Discovery grants. AS is supported by a SISSA Excellence Grant, and acknowledges partial support from the INFN-INDARK initiative. LP thanks Institute of Cosmology and Gravitation at University of Portsmouth for their hospitality. AH acknowledges useful discussions with Eric Linder.

\end{document}